\documentclass[prl,superscriptaddress,footinbib,twocolumn,showpacs,preprintnumbers,amsmath,amssymb,longbibliography,reprint]{revtex4-1}
\pdfoutput=1
\usepackage{subfigure}
\usepackage{dcolumn}
\usepackage{bm}
\usepackage{graphicx}
\usepackage{epsfig}
\usepackage{epstopdf}
\usepackage{multirow}
\usepackage{xspace}
\usepackage{array}
\usepackage{natbib}

\usepackage{atlasphysics}

\newcommand{\sch}   {\ensuremath{\tilde{c}}\xspace}

\newcommand{\neut}  {\tilde{\chi}^{0}_{1}}  
 
\def\etmiss{\ensuremath{E_{\mathrm{T}}^{\mathrm{miss}}}\xspace}

\def\dphimin{\ensuremath{\Delta\phi_{\mathrm{min}}}}
\def\DphiMin3{\ensuremath{\Delta\phi_{\mathrm{min}}(\ptmissvec, \text{3 jets})}}

\def\cjets{$c$-jets\xspace}
\def\cjet{$c$-jet\xspace}

\def\ctagged{$c$-tagged\xspace}

\def\Wjets{$W$+jets\xspace}
\def\Zjets{$Z$+jets\xspace}

\renewcommand{\ttbar} {\ensuremath{t\bar{t}}\xspace}

\renewcommand{\met} {\ensuremath{E_{\mathrm{T}}^{\mathrm{miss}}}\xspace}
\renewcommand{\pt} {\ensuremath{p_\mathrm{T}}\xspace}
\newcommand{\mct}{\ensuremath{m_\mathrm{CT}}\xspace}

\newcommand{\lumi}{20.3}
\newcommand\totallumi{20.3~${\rm fb}^{-1}$}

\def\met{\ensuremath{E_{\mathrm{T}}^{\mathrm{miss}}}}
\def\ptmissvec{\ensuremath{{\vec p_{\mathrm{T}}^{\mathrm{\hspace{1mm}miss}}}}\ }
\def\etmissvec{\ensuremath{{\vec E_{\mathrm{T}}^{\mathrm{miss}}}}\ }
\def\lsim{\mathrel{\rlap{\lower4pt\hbox{\hskip1pt$\sim$}}
    \raise1pt\hbox{$<$}}}                
\def\gsim{\mathrel{\rlap{\lower4pt\hbox{\hskip1pt$\sim$}}
    \raise1pt\hbox{$>$}}}                

\newcommand{\flavour}{flavor\xspace}
\newcommand{\flavours}{flavors\xspace}
\newcommand{\flavoured}{flavored\xspace}
\newcommand{\disfavour}{disfavor\xspace}

\newcommand{\twojTitle}{Search for Scalar Charm Quark Pair Production in $pp$ Collisions at $\sqrt{s}=$~8\,TeV with the ATLAS Detector}

\newcommand\twojabstracttext{The results of a dedicated search 
for pair production of scalar partners of charm quarks are reported. 
The search is based on an integrated luminosity of 
\totallumi~of $pp$ collisions at $\sqrt{s}=8$\,TeV recorded with
the ATLAS detector at the LHC. 
The search is performed using events with large missing transverse momentum 
and at least two jets, where the two leading jets are each tagged as originating from $c$ quarks. 
Events containing isolated electrons or muons are vetoed.
In an $R$-parity-conserving minimal supersymmetric scenario
in which a single scalar-charm state is kinematically accessible, 
and where it decays exclusively into a charm quark 
and a neutralino, 95\% confidence-level upper limits are obtained in the scalar-charm--neutralino 
mass plane such that, for neutralino masses below 200\,GeV, scalar-charm masses up to 
490\,GeV are excluded.}

\usepackage{preprintcover}
\usepackage{hyperref}
\hypersetup{colorlinks, allcolors=blue}
\PreprintCoverPaperTitle{\twojTitle}
\PreprintIdNumber{CERN-PH-EP-2014-292}
\PreprintCoverAbstract{\twojabstracttext}
\PreprintJournalName{Phys. Rev. Lett.}

\begin{document}
\title{\twojTitle}
\author{The ATLAS Collaboration}

\begin{abstract} 
\twojabstracttext
\end{abstract}
\pacs{14.80.Ly,13.85.Rm}

\maketitle
\normalsize

\newcommand{\ourplotsize}{0.75\columnwidth}
\newcommand{\ourbigplotsize}{0.88\columnwidth}
\newcommand{\ourleft}{top}
\newcommand{\ourright}{bottom}
\newcommand{\OurLetter}{Letter}
\newcommand{\ourfootnotespace}{~}

Supersymmetry 
(SUSY)~\cite{Miyazawa:1966mfa,Ramond:1971gb,Golfand:1971iw,Neveu:1971rx,Neveu:1971iv,Gervais:1971ji,Volkov:1973ix,Wess:1973kz,Wess:1974tw}
is a theory that extends the Standard Model (SM) and naturally resolves the hierarchy 
problem by introducing supersymmetric partners of the known bosons and 
fermions. In the framework of a generic $R$-parity-conserving minimal 
supersymmetric extension of the SM, the 
MSSM~\cite{Fayet:1976et,Fayet:1977yc,Farrar:1978xj,Fayet:1979sa,Dimopoulos:1981zb}, 
SUSY particles are produced in pairs and the lightest supersymmetric particle 
(LSP) is stable, providing a possible candidate for dark matter. In a large 
variety of models, the LSP is the lightest neutralino, $\ninoone$. 

The scalar partners (squarks) of various \flavours of quarks may, rather generally, 
have different masses despite constraints on quark \flavour mixing~\cite{Mahbubani:2012qq}.
Recent searches \disfavour low-mass top squarks (stops), sbottoms, 
and gluinos,
so direct scalar-charm (\sch) pair production could be the only squark production process accessible at the LHC.
Searches for \sch{} states provide not only a possible supersymmetry discovery mode 
but also the potential to probe the \flavour structure of the underlying theory.

Since no dedicated search for \sch{} has previously been performed, 
the best existing lower limits on \sch{} masses are obtained
from searches for generic squark and gluino production at the LHC~\cite{Aad:2014wea,Chatrchyan:2014lfa}, 
and from the reinterpretation of 
LHC searches~\cite{Aad:2014nra} for direct pair production 
of the scalar partner of the top quark followed by decays $\stopone \rightarrow c + \ninoone$.
The top squark searches have a final state similar to that expected for scalar charm quarks, 
but are optimized for small $m_{\tilde{t}} - m_{\ninoone}$ mass differences, 
and so have good sensitivity to the scalar charm quark only when $m_{\sch} - m_{\ninoone} \lsim m_W$.

In this \OurLetter{}, a dedicated search for direct \sch{} pair production is presented. The scalar charm quark is assumed to decay dominantly or exclusively via $\sch \rightarrow c + \ninoone$. 
The expected signal is therefore characterized by the presence of two jets originating from the hadronization of the $c$~quarks, accompanied by missing transverse momentum ($\etmiss$) resulting from the undetected neutralinos.

The ATLAS detector is described in detail elsewhere~\cite{DetectorPaper:2008}.
This search uses $pp$ collision data at a center-of-mass energy of 8\,\TeV{} 
recorded during 2012 at the LHC.
After the application of beam, detector and data quality requirements, 
the data set corresponds to a total integrated luminosity of 20.3~fb$^{-1}$ with a 
$2.8$\% uncertainty, using the methods of Ref.~\cite{Aad:2013ucp}.

The data are selected with a three-level trigger system that required a high transverse
momentum (\pt) jet and $\etmiss$~\cite{Aad:2012xs}.
While events containing charged leptons (electrons or muons) in the search region are vetoed, single-lepton triggers are used for control regions.
Events are required to have a reconstructed primary 
vertex consistent with the beam positions, 
and to meet basic quality criteria that suppress detector noise and noncollision 
backgrounds~\cite{Aad:2011JetEnergy}. 
Jets are reconstructed from three-dimensional topological calorimeter 
energy clusters by using the anti-$k_t$ jet algorithm~\cite{Cacciari:2008gp, Cacciari200657} 
with a radius parameter of 0.4. The measured jet energy is corrected 
for inhomogeneities and for the noncompensating response of the calorimeter 
by using $\pt$- and $\eta$-dependent\ourfootnotespace\footnote{ATLAS
  uses a coordinate system with its origin
  at the nominal interaction point (IP) in the center of the detector
  and the $z$-axis along the beam pipe.
  Cylindrical coordinates $(r,\phi)$ are used in the
  transverse plane, $\phi$ being the azimuthal angle around the beam
  pipe. The pseudorapidity is defined in terms of the polar angle
  $\theta$ as $\eta=-\ln\tan(\theta/2)$, while $\Delta R \equiv [ (\Delta\eta)^2 + (\Delta \phi)^2 ]^{1/2}$.}
 correction factors~\cite{JES}. The impact of multiple overlapping $pp$ interactions (pileup) is accounted for using a technique, based on jet areas, that provides an event-by-event and jet-by-jet correction~\cite{jetAreas}. 
Only jet candidates with $\pt > 20$\,GeV within $|\eta| < 2.8$ are retained. 

Electron candidates are required to have $\pt > 7$\,GeV, $|\eta| < 2.47$ 
and to satisfy ``medium'' selection criteria~\cite{Aad:2011Electrons}. 
Muon candidates are required to have $\pt > 6$\,GeV, $|\eta| < 2.4$ and are 
identified by matching an extrapolated inner-detector track to one or more 
track segments in the muon spectrometer~\cite{Aad:2014rra}. 
When defining lepton control regions, muons and electrons must meet additional ``tight'' selection criteria~\cite{Aad:2014rra,electronPerf},
and must satisfy track and calorimeter isolation criteria similar to those in Ref.~\cite{Aad:2013wta}.

Following this object reconstruction, overlaps between jet candidates and electrons or muons are resolved. Any jet within a distance $\Delta R $ = 0.2 of a medium quality electron candidate is discarded. Any remaining lepton within $\Delta R=$ 0.4 of a jet is discarded. Remaining muons must have longitudinal and transverse impact parameters within 1~mm and 0.2~mm of the primary vertex, respectively.

The calculation of $\etmiss$ is based on the vector sum 
of the calibrated $\pt$ of reconstructed jets (with $\pt > 20 \GeV$ and $|\eta| < 4.5$), 
electrons, muons and photons, and the calorimeter energy clusters 
not belonging to these reconstructed objects~\cite{ATLAS-MET-OLD}. 

Jets containing $c$-\flavoured hadrons without $b$-\flavoured parent hadrons
are identified using an algorithm, optimized for charm tagging, based on a 
neural network that exploits both impact parameter and secondary vertex 
information and with a $B$ to $D$ decay chain vertex fitter~\cite{JetFitterCharm}. 
This algorithm achieves a tagging efficiency of 19\% (13\%, 0.5\%) for $c$-jets ($b$-jets, light-\flavour or gluon jets)
in $t\bar{t}$ events.
The efficiency for tagging $b$-jets is 
determined from measurements of dileptonic $t\bar{t}$ events~\cite{ATLAS:BjetCalib}. 
The $c$-jet tagging efficiency and its uncertainty 
have been calibrated in inclusive jet events 
over a range of \pt{} using jets from collision 
data containing $D^*$ mesons 
~\cite{ATLAS:CLjetCalib}.
Jets can be $c$-tagged only within the acceptance of the inner detector ($|\eta|<2.5$), 
so only such central jets are retained after the above selection. 

Events are then required to have $\etmiss > 150$\,GeV and one jet with
$\pt>130$\,GeV to ensure full trigger efficiency, as well as a second jet with $\pt > 100$\,GeV. 
The two highest-\pt jets are required to be $c$ tagged.
The multijet background contribution with large \met, caused by 
mismeasurement of jet energies in the calorimeters or by neutrino 
production in heavy-quark decays, is suppressed by requiring a minimum 
azimuthal separation (\dphimin{}) of 0.4 between the $\etmiss$ direction 
and any of the three leading jets. 
To reduce the effect of pileup, the third jet is exempted from this requirement if it has $\pt < 50$\,GeV, $\left|\eta\right|<2.4$ and less than half of the sum of its track \pt is associated with tracks matched to the primary vertex.
In addition, the ratio of \met~to the 
scalar sum of the transverse momenta of the two leading jets is required to be above one-third.   
Events containing residual electron or muon candidates are vetoed in order to reduce electroweak backgrounds.

After these requirements, the main SM processes contributing to the background are top quark pair and single top production, together referred to as top production, as well as associated production of $W/Z$ bosons with light- and heavy-\flavour jets, referred to as \Wjets and \Zjets. A selection based on the boost-corrected contransverse mass \mct~\cite{Polesello:2009rn}  
is employed to further discriminate scalar-charm pair from top production. 
For two identical decays of heavy particles into two visible particles $v_{1}$ 
and $v_{2}$, and into invisible particles, the contransverse mass~\cite{Tovey:2008ui} 
is defined as 
$\{[ E_{\rm T}(v_{1}) + E_{\rm T}(v_{2}) ]^2 -  [  {\bf p_T}(v_{1}) - {\bf p_T}(v_{2}) ]^2\}^{1/2}$. 
The boost correction preserves the expected endpoint in the distribution against boosts caused by initial-state radiation. 
In the case of scalar-charm pair production with $\sch \rightarrow c+\neut$, \mct~is 
expected to have an endpoint at $(m_{\sch}^2-m_{\neut}^2)/m_{\sch}$. For \ttbar production, if both $b$-jets are mistagged as \cjets, the \mct\ built using those two jets is expected to have a kinematic endpoint at 135\,GeV.

To maximize the sensitivity across the $\sch$--$\neut$ mass plane, three overlapping signal regions (SR) are defined: \mct$\,> 150$, 200, and 250\,GeV. The remaining 
\ttbar background after the \mct\ requirement mostly comprises events with one true \cjet from a $W$ decay and a mistagged $b$-jet from a top quark decay. Events in which 
a $Z$ boson is produced in association with heavy-\flavour jets where the $Z$ boson decays into $\nu\bar{\nu}$ also enter the high-\mct\ regions. The heavy-\flavour jets often originate 
from a gluon splitting, $g \rightarrow c\bar{c}$, which can lead to a small angular separation between the resulting \cjets and therefore a small invariant mass $m_{cc}$. The remaining 
\ttbar{} background is also concentrated at low $m_{cc}$. Consequently, a final requirement selects events for which the invariant mass of the two $c$-tagged jets is larger than 200\,GeV.

Simulated-event samples are used to aid the description of the background and to model the SUSY signal. 
Top quark pair and single top production in the $s$ and $Wt$ channels are simulated with {\sc powheg-1.0 (r2092)}~\cite{powheg}, while the $t$ channel single top production is simulated using {\sc AcerMC 3.8}~\cite{Kersevan:2004yg}. A top quark mass of 172.5\,GeV is used. The parton shower, fragmentation, and hadronization are performed with {\sc pythia-6.426}~\cite{pythia}.
Samples of \Wjets, \Zjets, and dibosons ($WW$, $WZ$, $ZZ$) with light and heavy \flavour jets are generated with {\sc sherpa 1.4}~\cite{sherpa}, assuming massive $b$/$c$ quarks. 
Samples of $Z \ttbar$ and $W \ttbar$ are generated with {\sc MadGraph-5.1.3.33}~\cite{madgraph} interfaced to {\sc pythia-6.426}.
The signal samples are generated for a simplified SUSY model 
with only a single $\sch$ state kinematically accessible, 
and with BR($\sch \rightarrow c+\neut$)=100\%, 
using {\sc MadGraph-5.1.5.11} interfaced to {\sc pythia-6.427} for the parton shower, fragmentation, and hadronization. 
Signal cross sections are calculated to next-to-leading order in the strong coupling constant, adding the resummation of soft gluon emission at next-to-leading-logarithm accuracy (NLO+NLL)~\cite{Beenakker:1997ut,Beenakker:2010nq,Beenakker:2011fu}. The uncertainty on each nominal cross section is defined by an envelope of predictions using different PDF sets and factorization and renormalization scales, as described in Ref.~\cite{Kramer:2012bx}. 
The Monte Carlo (MC) samples are processed through a detector simulation~\cite{atlassimulation} based on {\sc geant4}~\cite{geant4}. The effects of pileup are included in the simulation. Efficiency corrections derived from the data are applied to the simulation to correct for lepton efficiency as well as the tagging and mistagging rates.

The main SM process contributing to the background after all signal region selections is \Zjets, followed by \Wjets, top quark pair, and single top production. Most \ttbar events contributing are $\ttbar \rightarrow bb\ell\nu qq$ events, in which either a $\tau$ lepton decays hadronically, or an $e$ or $\mu$ is out of the geometric acceptance or not reconstructed or identified.
Contributions from multijet, diboson, and associated production of \ttbar with $W,\ Z$ are subdominant. Noncollision backgrounds are found to be negligible. 

The estimation of the main background processes is carried out by defining a set of three data control regions (CR) that do not overlap with each other or with the signal regions. The CRs are kinematically close to the SRs and each of them is enhanced in one or two of the backgrounds that is dominant in the SRs, while having low expected signal contamination (less than $1\%$). A statistical model is set up in which the background expectation in the CRs and SRs depends on several parameters of interest: the normalizations of the dominant backgrounds, top (\ttbar + single top), \Zjets and \Wjets, as well as on nuisance parameters including the effect of uncertainties on the jet energy scale (JES) and resolution, calorimeter resolution for energy clusters not associated with any physics objects, energy scale and resolution of electrons and muons, $c$-tagging and mistagging rates, pileup, and luminosity. 
A profile likelihood fit of the background expectation to the data is performed simultaneously in all CRs~\cite{Baak:2014wma}, and from it the background normalizations are extracted. 
The normalization factors, which are consistent with unity within uncertainties, are then applied to the MC expectation in the signal regions.

The first control region is populated largely by \ttbar and \Wjets{}. It contains events
with exactly one isolated electron or muon with \pt above 50\,GeV. 
The leading two jets, with \pt$\, >$~130 and 50\,GeV respectively, must be $c$-tagged. 
To select events containing $W\rightarrow \ell \nu$, the transverse mass of the ($\ell, \etmiss$) system is required to be between 40 and 100\,GeV. The upper bound reduces possible signal contamination from SUSY models that produce leptons in cascade decays.
Finally, it is required that  $\etmiss >~100$\,GeV and $\mct > 150$\,GeV. 
The second control region is populated by $Z\rightarrow\ell^+\ell^-$ events with two opposite-sign, same-\flavour leptons, where the minimum \pt requirement is 70\,GeV for the leading lepton and 7(6)\,GeV for the subleading electron (muon). The transverse momenta of the leptons are added vectorially to the \etmissvec~to mimic the $Z\rightarrow \nu \bar{\nu}$ decay, and the modulus of the resulting two-vector is required to be larger than 100\,GeV. The leading two jets are required to be \ctagged and their \pt must each be above $50$\,GeV. The invariant mass $m_{\ell \ell}$ of the two leptons is required to be between 75 and 105\,GeV ($Z$-mass interval). 
A third control region, populated almost exclusively by dileptonic \ttbar events, contains events with two opposite-sign, different-\flavour leptons, where the leading lepton has $\pt > 25$\,GeV and the subleading lepton \pt is above 7(6) GeV for electrons (muons). It is required that $\etmiss >50$\,GeV and $m_{\ell \ell}> 50$\,GeV. The leading two jets are required to be \ctagged and have $\pt > 50$\,GeV. 
In all CRs, events with additional lepton candidates beyond the required number of signal leptons are vetoed using the same lepton requirements used to veto events in the SRs.

The subdominant background contributions from dibosons, $Z\ttbar$ and $W\ttbar$ are estimated by MC simulation. 
Finally, the residual multijet background is estimated using a data-driven technique based on the smearing of jets in a low-\etmiss~data sample with jet response functions~\cite{0leptonPaper}.

The experimental and theoretical uncertainties affecting the main backgrounds are correlated between control and signal regions, and the data observed in control regions constrain the uncertainties on the expected yields in the signal regions.
The residual uncertainty due to the theoretical modeling of the top-production background is about 7\%. 
It is evaluated using additional MC samples generated with {\sc AcerMC} (where initial- and final-state 
radiation parameters are varied) an alternative fragmentation model ({\sc herwig}), an alternative generator ({\sc mc@nlo}), 
and 
by using diagram subtraction rather than diagram removal to account for the interference between \ttbar and single top $Wt$-channel production \cite{singleTopDRDS}.
After the fit, the residual uncertainties on the \Wjets~and~\Zjets~theoretical modeling account for less than 20\% of the total uncertainty. 
The dominant contributions to the residual uncertainty on the total background are from 
$c$-tagging ($\sim$20\%), 
normalization uncertainties related to the numbers of events in the CRs (10\%--20\%), 
and JES ($\sim$10\%).

For the SUSY signal processes, theoretical uncertainties on the cross section due to the choice of renormalization and factorization scales and from PDFs are found to be between 14\% and 16\% for \sch{} masses between 100 and 550\,GeV. 
Prior to the fit, the detector-related uncertainties with largest impact on the signal event yields are those for $c$-tagging (typically 15\%--30\%) and JES (typically 10\%--30\%).

\begin{table*}[htp]
  \small
  \newcommand\bracketsize\small
  \renewcommand\arraystretch{1.1}
  \setlength\tabcolsep{3em}
  \centering

\begin{tabular}{l c c c}
\hline\hline
\mct{} (GeV)  & $>$150 & $>$200 & $>$250 \\
\hline 
Top       & $7.4 \pm 2.7 \hspace{1.5em} (7.1)$ & $3.9 \pm 1.6 \hspace{1.5em} (3.7)$ & $1.6 \pm 0.7 \hspace{1.5em} (1.5)$ \\
$Z$+jets  & $14 \pm 3 \hspace{1.5em}\hspace{0.78em} (13)$ & $7.7 \pm 1.7 \hspace{1.5em} (7.0)$ & $4.3 \pm 1.2 \hspace{1.5em} (3.9)$ \\
$W$+jets  & $7.2 \pm 4.5 \hspace{1.5em} (7.4)$ & $4.1 \pm 2.6 \hspace{1.5em} (4.2)$ & $1.9 \pm 1.2 \hspace{1.5em} (1.9)$ \\
Multijets & $0.3 \pm 0.3$ &  $0.2 \pm 0.2 $ &  $0.05 \pm 0.05$ \\
Others    & $0.5 \pm 0.3$ & $0.4 \pm 0.3$ & $0.4 \pm 0.3$ \\ [2pt]
{\bf Total} & {\bf $30 \pm 6$ } & {\bf $16 \pm 3$ } & {\bf $8.2 \pm 1.9$ } \\
{\bf Data}  & {\bf 19 }& {\bf 11 }& {\bf 4 }\\
\hline\hline
\end{tabular}
  \caption{Expected and observed number of events for 
    an integrated luminosity of \lumi\,fb$^{-1}$ at $\sqrt{s}=8$\,TeV. 
    Top, \Zjets and \Wjets contributions are estimated using the fit described in the text. For comparison, the numbers obtained using MC simulations only are shown in parentheses.
    The row labeled ``Others'' reports subdominant electroweak backgrounds estimated from MC simulations. 
    The total uncertainties are also reported.} 
  \label{tab:0lepFinalResultsSR_2jX_ROTATED}
\end{table*}

\begin{figure}[!h]
\centering
 \includegraphics[width=\ourplotsize]{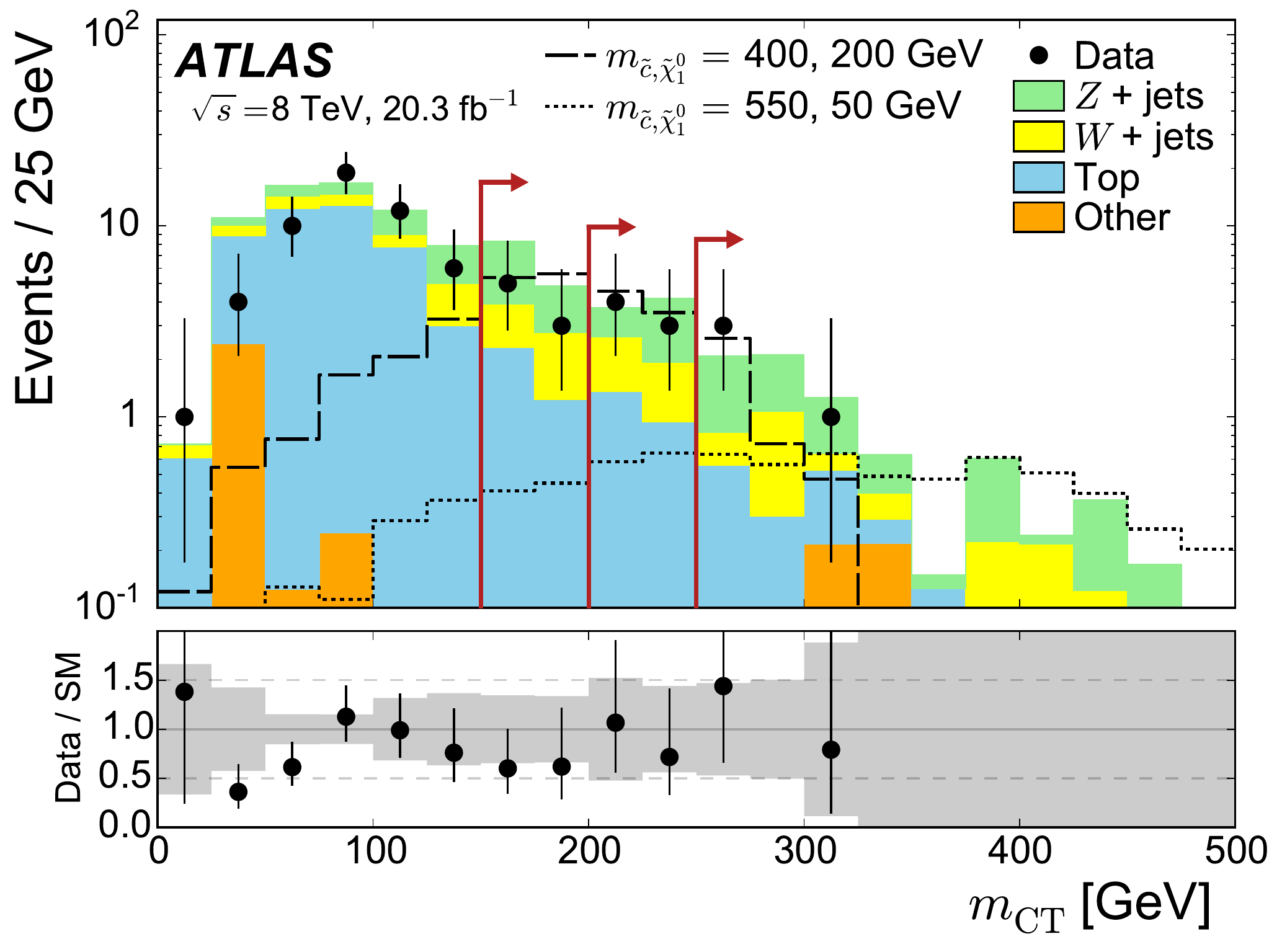}
 \includegraphics[width=\ourplotsize]{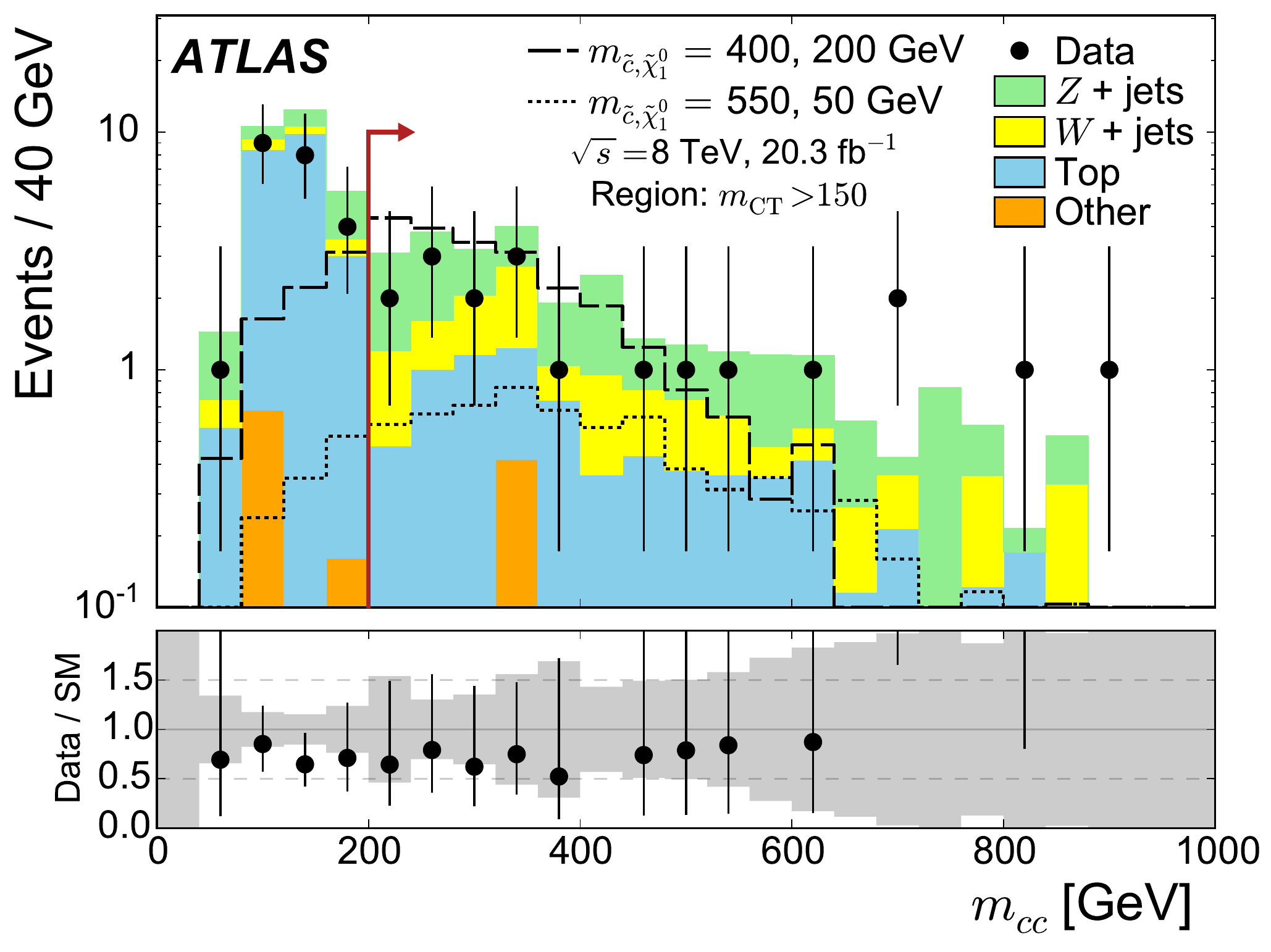}
 \caption{Distributions of \mct{} (\ourleft) and $m_{cc}$ (\ourright), and their corresponding SM predictions. Signal region selections ($\mct>150\,\GeV$ for the $m_{cc}$ distribution) are applied, other than for the variable plotted. Arrows indicate the SR requirements on \mct{} and $m_{cc}$. In the ratio plots, the grey bands correspond to the combined MC statistical and experimental systematic uncertainty.}
 \label{fig:res_plots}
\end{figure}

Table~\ref{tab:0lepFinalResultsSR_2jX_ROTATED} reports the observed number of events and the SM predictions 
for each SR. 
The data are found to be below the SM background expectations, 
but consistent with them given the uncertainties.  
Figure~\ref{fig:res_plots} shows the measured \mct~and $m_{cc}$~distributions in the $\mct>150$\,GeV region compared to the SM predictions. Monte Carlo 
estimates are shown after the normalizations 
extracted from the profile likelihood fit are applied. For illustrative purposes, the distributions expected for the simplified model with $(\sch,~\neut)$ masses 
of (400,\,200)\,GeV and (550,\,50)\,GeV are also shown.

\begin{figure}[h!]
  \begin{center}
   \includegraphics[width=0.86\columnwidth,trim=0 40 0 0]{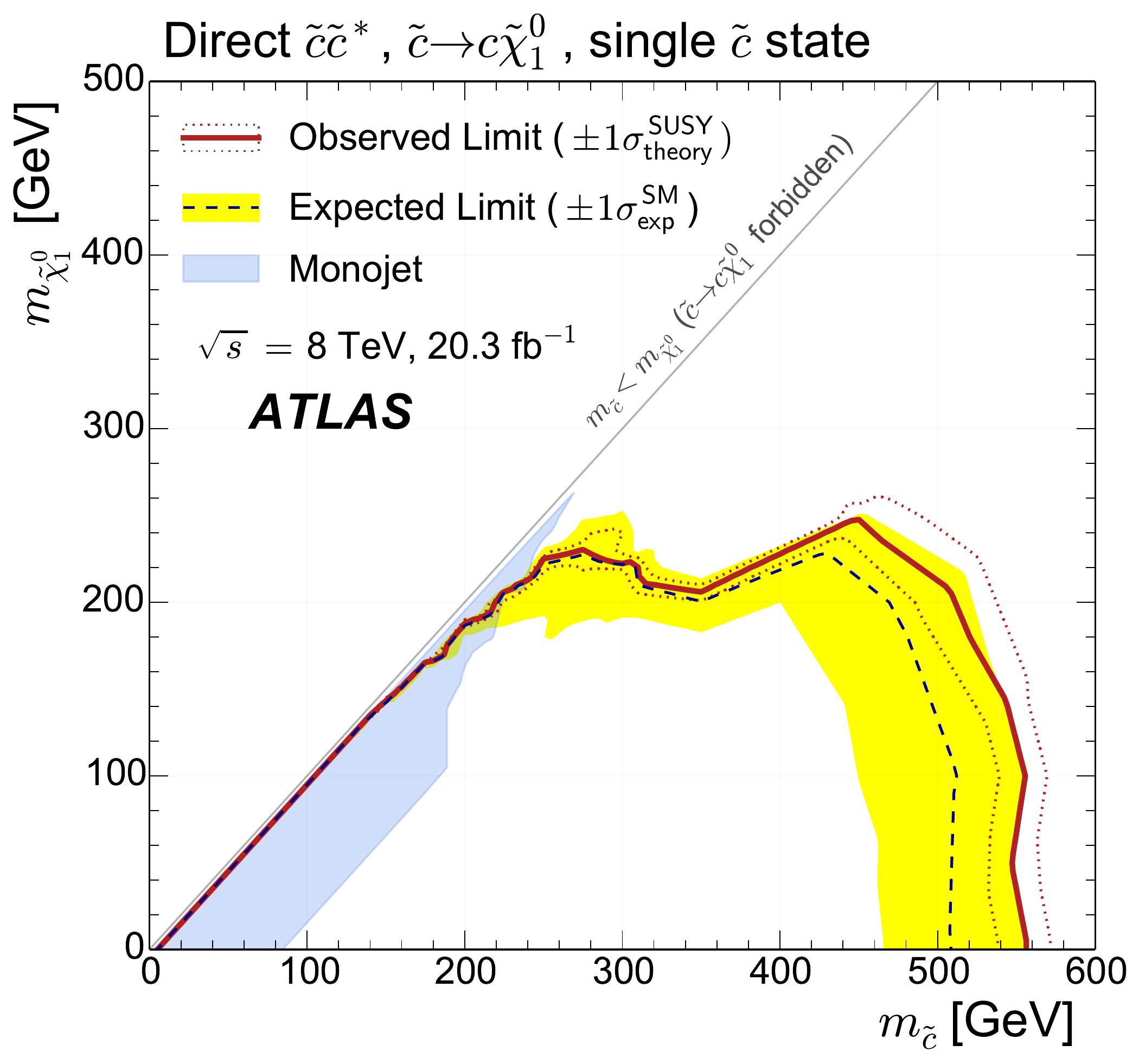}
  \end{center}
  \caption{Exclusion limits at 95\% C.L. in the $\sch$--$\neut$ mass plane. The observed (solid red line) and expected (dashed blue line) limits include all 
uncertainties except for the theoretical signal cross-section uncertainty (PDF and scale). The band 
around the expected limits show $\pm 1 \sigma$ uncertainties. The dotted lines around the observed limits represent the results obtained when moving the nominal signal cross 
section up or down by the $\pm 1 \sigma$ theoretical uncertainty.}
\label{fig:excl_scharm_LSP}
\end{figure}
Since no significant excesses are observed, the results are translated into 95\% confidence-level (C.L.) upper limits on contributions from non-SM processes using 
the CL$^{}_{\rm s}$ prescription~\cite{profilelike}. 
Figure~\ref{fig:excl_scharm_LSP} shows the observed and expected exclusion limits at 95\% C.L. on the $\sch$--$\neut$ mass plane, assuming a single accessible $\sch$ 
particle with BR($\sch \rightarrow c+\neut)=100\%$. 
The SR with the best expected sensitivity at each point in the plot is adopted as the nominal result. 
In the region where the $c$-tagged analysis of the ATLAS $\tilde{t}\rightarrow c + \neut$ search~\cite{Aad:2014nra} 
provides a stronger expected limit, i.e. for $m_{\sch}-m_{\neut} \lsim m_W$, that result is used. 
The region excluded by the ATLAS monojet search described in Ref.~\cite{Aad:2014nra} is shown separately as a grey shaded area. 
Systematic uncertainties, other than in the \sch{} pair-production cross section, 
are treated as nuisance parameters and correlated when appropriate. 
For the SUSY scenario considered, the upper limit at 95\% C.L. on the scalar-charm mass 
obtained in the most conservative cross-section hypothesis
is 540\,GeV for $m^{}_{\neut} = 0$ (increasing to 555\,GeV for the central estimate of the signal cross section). 
Neutralino masses up to 200\,GeV are similarly excluded for $m_{\sch} < 490$\,GeV.  
This significantly extends the results of previous \flavour-blind analyses~\cite{Aad:2014wea,Chatrchyan:2014lfa}, which provide no exclusion for $m_{\neut} > 160$\,GeV,
nor for single light squarks with masses above 440\,GeV.
The signal regions are used to set limits on the effective cross sections $\sigma^{}_{\rm vis}$ of any non-SM processes, including the effects of branching ratios, experimental acceptance, and efficiency, 
neglecting any possible contamination in the control regions.
Values of $\sigma^{}_{\rm vis}$ larger than 0.44\,fb, 0.36\,fb, and 0.23\,fb are excluded at 95\% C.L. 
for \mct{} greater than 150, 200, and 250\,GeV respectively.

In summary, this \OurLetter{} reports results of a search for scalar-charm pair production in 8\,TeV $pp$ collisions at the LHC, based on \lumi\,fb$^{-1}$ of ATLAS data. 
The selected events have large \met~and two \ctagged jets. The results are in agreement with SM predictions for backgrounds and 
translate into 95\% C.L. upper limits on scalar-charm and neutralino masses in a simplified model with a single accessible \sch{} state for which the exclusive decay $\sch \rightarrow c+\neut$ is 
assumed. For neutralino masses below 200\,GeV, scalar-charm masses up to 490\,GeV are excluded, significantly extending previous limits.

\vspace{1em}

We thank CERN for the very successful operation of the LHC, as well as the
support staff from our institutions without whom ATLAS could not be
operated efficiently.

We acknowledge the support of ANPCyT, Argentina; YerPhI, Armenia; ARC,
Australia; BMWFW and FWF, Austria; ANAS, Azerbaijan; SSTC, Belarus; CNPq and FAPESP,
Brazil; NSERC, NRC and CFI, Canada; CERN; CONICYT, Chile; CAS, MOST and NSFC,
China; COLCIENCIAS, Colombia; MSMT CR, MPO CR and VSC CR, Czech Republic;
DNRF, DNSRC and Lundbeck Foundation, Denmark; EPLANET, ERC and NSRF, European Union;
IN2P3-CNRS, CEA-DSM/IRFU, France; GNSF, Georgia; BMBF, DFG, HGF, MPG and AvH
Foundation, Germany; GSRT and NSRF, Greece; ISF, MINERVA, GIF, I-CORE and Benoziyo Center,
Israel; INFN, Italy; MEXT and JSPS, Japan; CNRST, Morocco; FOM and NWO,
Netherlands; BRF and RCN, Norway; MNiSW and NCN, Poland; GRICES and FCT, Portugal; MNE/IFA, Romania; MES of Russia and ROSATOM, Russian Federation; JINR; MSTD,
Serbia; MSSR, Slovakia; ARRS and MIZ\v{S}, Slovenia; DST/NRF, South Africa;
MINECO, Spain; SRC and Wallenberg Foundation, Sweden; SER, SNSF and Cantons of
Bern and Geneva, Switzerland; NSC, Taiwan; TAEK, Turkey; STFC, the Royal
Society and Leverhulme Trust, United Kingdom; DOE and NSF, United States of
America.

The crucial computing support from all WLCG partners is acknowledged
gratefully, in particular from CERN and the ATLAS Tier-1 facilities at
TRIUMF (Canada), NDGF (Denmark, Norway, Sweden), CC-IN2P3 (France),
KIT/GridKA (Germany), INFN-CNAF (Italy), NL-T1 (Netherlands), PIC (Spain),
ASGC (Taiwan), RAL (UK) and BNL (USA) and in the Tier-2 facilities
worldwide.

\clearpage 

\clearpage 

\onecolumngrid
\begin{flushleft}
{\Large The ATLAS Collaboration}

\bigskip

G.~Aad$^{\rm 85}$,
B.~Abbott$^{\rm 113}$,
J.~Abdallah$^{\rm 152}$,
S.~Abdel~Khalek$^{\rm 117}$,
O.~Abdinov$^{\rm 11}$,
R.~Aben$^{\rm 107}$,
B.~Abi$^{\rm 114}$,
M.~Abolins$^{\rm 90}$,
O.S.~AbouZeid$^{\rm 159}$,
H.~Abramowicz$^{\rm 154}$
% ...
\bigskip
\\
$^{1}$ Department of Physics, University of Adelaide, Adelaide, Australia\\
% ...
$^{11}$ Institute of Physics, Azerbaijan Academy of Sciences, Baku, Azerbaijan\\
% ...
$^{85}$ CPPM, Aix-Marseille Universit{\'e} and CNRS/IN2P3, Marseille, France\\
% ...
$^{90}$ Department of Physics and Astronomy, Michigan State University, East Lansing MI, United States of America\\
% ...
$^{107}$ Nikhef National Institute for Subatomic Physics and University of Amsterdam, Amsterdam, Netherlands\\
% ...
$^{113}$ Homer L. Dodge Department of Physics and Astronomy, University of Oklahoma, Norman OK, United States of America\\
$^{114}$ Department of Physics, Oklahoma State University, Stillwater OK, United States of America\\
% ...
$^{117}$ LAL, Universit{\'e} Paris-Sud and CNRS/IN2P3, Orsay, France\\
% ...
$^{152}$ Institute of Physics, Academia Sinica, Taipei, Taiwan\\
% ...
$^{154}$ Raymond and Beverly Sackler School of Physics and Astronomy, Tel Aviv University, Tel Aviv, Israel\\
% ...
$^{159}$ Department of Physics, University of Toronto, Toronto ON, Canada\\
% ...
\end{flushleft}


\end{document}